\documentclass[11pt]{article}
\usepackage{moriond,amssymb}

\def\be{\begin{equation}}
\def\ee{\end{equation}}
\def\ba{\begin{array}}
\def\ea{\end{array}}
\def\bt{\begin{table}[ht]}
\def\et{\end{table}}
\def\btab{\begin{tabular}}
\def\etab{\end{tabular}}
\def\bc{\begin{center}}
\def\ec{\end{center}}
\def\rep#1{{\bf #1}}
\def\ov#1{\overline{#1}}

\def\jrn#1#2#3#4{{#1} {\bf #2} (#4) #3}
\def\PRL{Phys. Rev. Lett.}
\def\PLB{Phys. Lett. B}
\def\PRD{Phys. Rev. D}

\begin{document}

\vspace*{0cm}

\title{Neutrino Mixings and Unified Theories}

\author{Fu-Sin Ling}

\address{Institute for Fundamental Theory\\
Department of Physics, University of Florida,\\ 
Gainesville, FL, 32611, USA\\}

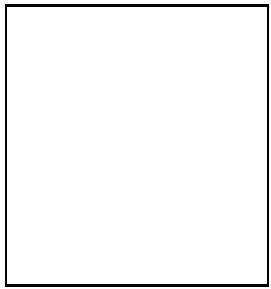
\maketitle

\begin{abstract}
The recent neutrino data converge towards a 3$\nu$ scheme
with two large and one small mixing angles.
Their implications for model-building are discussed.
Some possible components of a flavor symmetry over 
quarks and leptons are singled out.
We suggest that they can be naturally embedded in a unified 
structure with a replicated gauge group.
\end{abstract}

\section{Introduction}

The last five years have proven to be incredibly rich
in Neutrino Physics. Numerous experiments around the
globe enabled to turn neutrino oscillations from discovery 
into an almost complete determination of the physical parameters.
Noticeably, the confirmation of the original oscillation signals 
from solar and atmospheric neutrinos by terrestrial and man-made 
sources stands as a remarkable success~\cite{kamland,K2K} 
and leads to a consistent picture of the neutrino sector.

The scheme that emerges involves the three active neutrinos,
whose light mass eigenstates~\footnote{in the effective low-energy
description with Majorana masses} are related to the flavor eigenstates
by a mixing matrix $U_{MNS}$ (the Maki-Nakagawa-Sakata matrix)~\footnote{
in the charged leptons mass eigenstates basis} that contains two
large and one small mixing angles. 
The presence of a small mixing angle $\theta_{13}$ 
enables to write $U_{MNS}$ as
\be
U_{MNS} ~\simeq~ \left( 
\ba{ccc}
\cos \theta _\odot & \sin \theta _\odot & s_{13}e^{-i \delta} \\
-\sin \theta _\odot \cos \theta _\oplus & 
\cos \theta _\odot \cos \theta _\oplus & \sin \theta _\oplus \\
\sin \theta _\odot \sin \theta _\oplus & 
-\cos \theta _\odot \sin \theta _\oplus & \cos \theta _\oplus \\
\ea
\right)
\ee
where $\theta _\odot \simeq \pi/6$ and $\theta _\oplus \simeq \pi/4$
are the mixing angles measured by solar and atmospheric neutrinos
experiments respectively,
$s_{13} = \sin^2 \theta _{13}$ and $\delta$ is a possible
CP violating phase. The constraint from the CHOOZ 
experiment~\cite{CHOOZ} translates into a bound for $\theta _{13}$
\be
\sin^2 \theta _{13} ~\lesssim~ 0.03
\ee
A mild hierarchy between $\Delta m^2_\odot$ for solar neutrinos 
and $\Delta m^2_\oplus$ for atmospheric neutrinos is also 
observed~\cite{SKatm,SKsol}, although the absolute 
scale for neutrinos masses is yet to be found.
Assuming the LMA solution for solar neutrinos (combined with the
KamLAND result), we have
\be
\Delta m^2_\odot ~\simeq~ 7 \cdot 10^{-5} \; {\rm eV}^2 \; ; \qquad
\Delta m^2_\oplus ~\simeq~ 3 \cdot 10^{-3} \; {\rm eV}^2 
\ee
so that
\be
\frac{\Delta m^2_\odot}{\Delta m^2_\oplus} ~\simeq~ 10^{-2}
\ee 

The experimental evidences that support this 
picture can be summarized as follows. 
The SNO neutral current data~\cite{SNO} proves beyond any doubt that
flavor conversion of solar neutrinos is taking place.
Moreover, it limits severely the amount of
neutrino disappearance in sterile flavors or in extra-dimensions.
The observed loss of reactor anti-neutrinos in KamLAND agrees with
the solar neutrinos data, and selects the LMA region as the unique
solution. Moreover, the possibility of CPT violation in the neutrino 
sector, although not completely ruled out, seems unlikely. 
Of course, the controversial claim of the LSND experiment~\cite{LSND} 
is quite problematic in this picture, because a $4 \nu$ scheme 
gives only a poor fit to the data. 
It will need the verdict of miniBoone, meanwhile, one solution is
simply to drop out the LSND result. 

It appears that the Yukawa structure in the neutrino sector looks 
very different from that of the quark sector, where steep mass 
hierarchies come along with small mixing angles.
From the model-building point of view, the latter situation
is natural, because a hierarchical structure in a mass matrix
will precisely yield hierarchical masses and small mixing angles.
For example, with two flavors, a mass matrix with the structure
(the ${\cal O}(1)$ prefactor coefficients are dropped)
\be
\left( 
\ba{cc}
\epsilon^2 & \epsilon \\
\epsilon & 1
\ea
\right)
\ee
with $\epsilon \ll 1$ gives two eigenmasses $m_1 \sim {\cal O}(1)$
and $m_2 \sim {\cal O}(\epsilon ^2)$, and a mixing angle
$\theta \sim {\cal O}(\epsilon)$.
On the contrary, hierarchical masses with large mixings lead 
to a mass matrix without hierarchical structure, but
with correlated entries. Again, with two flavors, 
if $m_1 \sim {\cal O}(1)$, $m_2 \sim {\cal O}(\epsilon^2)$
and $\theta = \pi /4$ we get a matrix of the form
\be
\left( 
\ba{cc}
1 & 1 \\
1 & 1
\ea
 \right)
\ee
whose determinant is $D \sim {\cal O}(\epsilon^2)$. 
Of course, the absolute mass scale of neutrinos is not known yet,
but naturalness arguments in unified theories favor hierarchical
masses over almost degenerate masses with hyperfine splittings.

The question is to understand how large mixing angles have to be 
treated in a model-building point of view.
In what follows, we analyze their relationships with hierarchical
mass patterns to derive possible hidden components of a unifying
structure. We then show how these components can be embedded in
a scheme with a replicated gauge group.

\section{Schemes with large neutrinos mixings}

We start by reconstructing the effective neutrino mass matrix,
assuming hierarchical masses, to obtain the following
hierarchical structure
\be
\label{hierstruct}
\left( 
\ba{ccc}
\epsilon & \epsilon & \epsilon \\
\epsilon & 1 & 1 \\
\epsilon & 1 & 1 \\
\ea
\right)
\ee
The parameter $\epsilon$ is set by the measured ratio 
of $\Delta m^2$, so $\epsilon^2 \sim 10^{-2}$.

It is worth pointing out that the hierarchy in the 
neutrino sector is very mild compared to that 
observed in the charged lepton or in the quark sectors. 
This observation, together with the appearance of large
mixing angles, supports the idea of anarchy as a hypothesis 
for the description of the neutrino sector.
A first task is to compare this possibility
to a hierarchical structure.

\subsection{Anarchy}

Anarchy is basically the idea that there is no fundamental
distinction between the three different flavors of neutrinos.
The neutrino mass matrix will therefore have no structure
\be
\left( 
\ba{ccc}
1 & 1 & 1 \\
1 & 1 & 1 \\
1 & 1 & 1 \\
\ea
\right)
\ee
and the observed values of mixing angles and $\Delta m^2$
ratio just reflect Nature's choice of the random ${\cal O}(1)$
prefactors that conspired to produce the observed data.

In this scenario, large mixing angles are expected to be generic,
so that the small value of $\theta _{13}$ puts a severe constraint.
Anarchical matrices have been analyzed using statistical 
tools~\cite{anan}, and have been shown to be consistent with 
data with a probability of about 12\%.

The anarchical hypothesis can also be embedded in a 
model~\cite{anarchy} based on an Abelian family 
symmetry commuting with $SU(5)$, which can reproduce the
observed hierarchies among quarks and leptons.  
This step towards a quark-lepton unification in the Yukawa
couplings finds roots in the mass relation $m_b \simeq m_\tau$.
However, the agreement in the measured masses between down-type 
quarks and charged leptons deteriorates for the lighter families. 
As argued in an earlier work~\cite{me1}, in the framework of 
Abelian family symmetries, neutrino data could also hint 
at some quark-lepton unification. 
In this context, it is natural to ask how large mixing angles 
can appear in the lepton sector while only small mixing angles
are observed in the quark sector. 
In the case of a family symmetry that commutes with $SU(5)$, 
the answer is that the charged leptons and the down-type quarks 
mass matrices are transposed of one another, and therefore, 
large mixing angles in the lepton sector correspond to 
hidden, unobservable rotations in the quark sector.

We also note that the anarchical scenario for neutrinos is a possible 
outcome in a unified scheme with a replicated gauge group~\cite{me2}. 

\subsection{Hierarchy}

If the neutrino mass matrix is given by Eq.~(\ref{hierstruct}),
the values of the mixing angles will also strongly depend on how 
the prefactor coefficients conspire.

In the generic case, {\it i.e.} when there is no cancellation
between these coefficients, the sub-determinant of the 2-3
block is $D_1 \sim {\cal O}(1)$, so we expect the following
behavior for eigenmasses and mixing angles,
\be
\mu _1 \sim {\cal O}(\epsilon), \;\;
\mu _{2,3} \sim {\cal O}(1); \qquad
\theta _\oplus \sim {\cal O}(1), \;\;
\theta _\odot, s_{13} \sim {\cal O}(\epsilon)
\ee
If the sub-determinant is suppressed, the solar mixing angle
gets enhanced. It becomes large, $\theta _\odot \sim {\cal O}(1)$,
for $D_1 \sim {\cal O}(\epsilon)$, and close to maximal
for $D_1 \sim {\cal O}(\epsilon^p)$, with $p>1$.

The sensitivity of the solar mixing angle to the value
of the sub-determinant implies that radiative corrections might
play an important role. When the neutrino masses are obtained by 
the see-saw mechanism, the corrections are indeed important,
and depend strongly on the masses of the right-handed 
neutrinos~\cite{rc}. On the other hand, the mixing angles
$\theta _\oplus$ and $\theta _{13}$ are more stable.

If radiative corrections are indeed important for the solar
mixing angle, low-energy and high-energy parameters will not 
be simply related. In particular, the correspondence will
presumably depend on the unknown masses of the right-handed neutrinos.
These difficulties can be avoided in several cases. 

For example, in the situation referred to as "right-handed
neutrino dominance~\cite{king}", only one right-handed neutrino
is responsible for the leading contributions to the low-energy
mass matrix obtained.  
As an illustration, we take two flavors and
\be
m_D ~=~ m \left( \ba{cc} a & b\\ c & d\\ \ea \right)\, , \qquad
M^{-1} ~=~ \frac{1}{M_0} \left( \ba{cc} 1 & 0\\ 0 & 0\\ \ea \right)
\ee 
This gives a low-energy neutrino mass matrix with a vanishing 
determinant
\be
m_\nu ~=~ m_D \, M^{-1} \, m_D^t ~=~ \frac{m^2}{M_0}
\left( \ba{cc} a^2 & ac\\ ac & c^2 \ea \right)
\ee
The other right-handed 
neutrinos give rise to sub-leading contributions.
As a result, the sub-determinant
can be naturally small, even if radiative corrections are added.
Therefore, a natural large solar mixing angle can be obtained
if right-handed neutrino dominance holds.
However, it is not clear that $\theta _\odot$
will be stable against radiative corrections, because its
precise value is a ratio of sub-leading contributions.

Right-handed neutrino dominance is however not a feature of models
with $U(1)$ family symmetries of the chiral type.
The basic reason is that the right-handed neutrino family charges
cancel in the see-saw. The structure of the neutrino matrix is 
given by the charges of the lepton doublet
\be
m_\nu ~\sim~ L \bar{N} \cdot \frac{1}{\bar{N} \bar{N}^t} 
\cdot (L \bar{N})^t ~\sim~ L L^t
\ee
and the value of the sub-determinant $D_1$ is not 
generically suppressed.

\subsection{Discrete symmetries}

Another way to avoid the sensitivity to radiative corrections is to
impose more symmetry. One can think of non-Abelian symmetries,
or discrete symmetries as additional components of the 
flavor structure. 

Discrete flavor symmetries in the neutrino mass matrix give the 
possibility to enforce special values of the mixing angles, as
it is suggested by the recent neutrino data.
For example, with two generations, a permutation symmetry $P_{12}$
between the two flavors automatically imposes a maximal 
mixing angle. With three generations, an approximate $P_{23}$
symmetry can give rise to a maximal atmospheric angle, and a large but
not maximal solar angle~\cite{me1}. 

An extension of a permutation symmetry to three flavors leads to
the finite non-Abelian group $S_3$ of permutations of three objects.  
It turns out~\cite{S3} that a tri-bimaximal neutrino mixing 
can be achieved if neutrinos belong to the (reducible) 
three dimensional representation of $S_3$.
The tri-bimaximal mixing corresponds to a bi-maximally mixed neutrino
$\nu _3$ and a tri-maximally mixed neutrino $\nu _2$, corresponding
to a maximal atmospheric angle $\theta _\oplus = 45^\circ$, and
a large solar angle of about $\theta _\odot \simeq 35^\circ$.
\be
U ~=~ \left( 
\ba{ccc}
\sqrt{\frac{2}{3}} & \frac{1}{\sqrt{3}} & 0 \\
-\frac{1}{\sqrt{6}} & \frac{1}{\sqrt{3}} & -\frac{1}{\sqrt{2}} \\
-\frac{1}{\sqrt{6}} & \frac{1}{\sqrt{3}} & \frac{1}{\sqrt{2}} \\
\ea
\right)
\ee

\section{Flavor symmetries revisited}

Because the Yukawa structure and the mixing angles are very
different for quarks and for leptons, it is difficult to embed 
naturally both sectors in a unified scheme. 
A hierarchical structure that arises from an Abelian family
symmetry can fit nicely the quark sector but
leaves the neutrino sector sensitive to order one parameters and
to possible radiative corrections. On the other hand, a discrete
symmetry that would explain the particular values of the neutrino 
mixings would require severe fine-tuning to accommodate the hierarchical
mass patterns for quarks. 

How can Abelian flavor symmetries and discrete flavor symmetries
emerge from a common scheme?
We contemplate the idea that both components emerge from
a replicated gauge structure of the type
\be
G_0 ~=~ G \times G \times G
\ee
The fundamental gauge group at the Planck scale contains a copy
of the same group $G$ for each generation.
Fermions for each family fall into representations of the
respective copy of $G$, but are singlets under the other copies.
By symmetry-breaking, the fundamental group is reduced to the
diagonal (family-blind) Standard Model gauge group
$SU(3) \times SU(2) \times U(1)_Y$.
In the chain of symmetry breakings, family-dependent Abelian symmetries 
are left over at some intermediate step. We assume that they are
responsible for the observed hierarchies of the Yukawa couplings. 
In this framework, the question of family hierarchy reduces 
to a search for plausible unifying groups and the way to break them. 
The patterns of the fermion masses and mixings are encoded in the 
underlying group structure and the breaking path~\cite{me2}.

Because the low-energy spectrum contains chiral fermions, a good
candidate for a replicated gauge group must be anomaly-free and
have complex representations that contain the known fermions.
This limits the freedom to the sequence
\be
E_6 ~\supset~ SO(10) ~\supset~ SU(5)
\ee

It turns out that phenomenologically relevant $U(1)$ 
symmetries can arise in the breaking
\be
\label{e6}
E_6 \times E_6 \times E_6 \; \longrightarrow \; SU(5)
\ee
We refer the interested reader to the paper~\cite{me2} for a detailed
analysis of the symmetry breaking and the way they can give rise
to family-dependent $U(1)$ symmetries.
Here, we would only like to point out that this construction
leads to Abelian flavor symmetries that are automatically
anomaly-free. Moreover, their form is not arbitrary, but is
reminiscent of the group structure and the breaking path,
although the (finite) number of such possible flavor symmetries
turns out to be quite large.

The breaking scheme of Eq.~(\ref{e6}) leads to a few viable
mass patterns from the point of view of phenomenology.
They are expressed by the family charges for the two irreducible
representations $\rep{10}$ and $\rep{\ov{5}}$ of $SU(5)$ 
(The family symmetry $Y_F$ commutes with $SU(5)$ because of the
breaking scheme that has been chosen). Three models are listed in
Table~\ref{3models}.
\bt
\bc
\btab{cc|cccc}
& & & $Y_F(\rep{10})$ & & $Y_F(\rep{\ov{5}})$ \\ &&&&&\\
\hline &&&&&\\
Model A & & & $(4,2,0)$ & & $(0,0,0)$ \\ &&&&&\\
Model B & & & $(3,2,0)$ & & $(2,0,0)$ \\ &&&&&\\
Model C & & & $(5,2,0)$ & & $(1,0,0)$ \\ &&&&&\\
\hline
\etab
\ec
\caption{\label{3models} Family charges for three models with
phenomenological relevance, obtained in the symmetry breaking
$E_6 \times E_6 \times E_6 ~\rightarrow~ SU(5)$.}
\et
All these models lead to the following relations, strongly
suggested by data,
\be
\frac{m_c}{m_t} ~\sim~ \left( \frac{m_s}{m_b} \right)^2
~\sim~ \left( \frac{m_\mu}{m_\tau} \right)^2
~\sim~ \lambda^4
\label{eq:constraint}
\ee
where the expansion parameter $\lambda$ has a value close to
that of the Cabibbo angle $\lambda _c \simeq 0.22$.
As a reward, the relations in Eq.~(\ref{eq:constraint})
automatically give rise to a large mixing angle for the 
atmospheric neutrinos!

As we can see, these models realize a compromise between the
strong hierarchy observed in the quark sector, and the mild 
hierarchy with large mixing angles in the neutrino sector.
In particular, the model A gives rise to an anarchical scenario
for the neutrinos. We suggest that a discrete flavor
symmetry plays a critical role in this last case.
In the replicated gauge structure, it is indeed natural to 
suppose that the full lagrangian at the Planck scale is 
invariant under the permutation 
group $S_3$ between the three families. 
In the quark sector, this symmetry is spontaneously
broken by the vacuum expectation value of the order parameter,
and the $U(1)$ flavor symmetry drives a hierarchical structure.
However, the flavor gauge structure being absent for neutrinos,
an exact $S_3$ symmetry can remain. If this is the case,
the low-energy neutrino mixing angles that are being measured
may be remnants of a replicated structure that holds at a very
early stage. A complete understanding of this picture is still 
in progress~\cite{proj}.

\section{Conclusions}

The advent of large mixing angles in the neutrino sector brings 
a new challenge for theorists. While hierarchical structures
that arise from $U(1)$ flavor symmetries are well suited to describe
the quark sector, their predictions for neutrinos become sensitive
to ${\cal O}(1)$ coefficients and radiative corrections.
Rather, a discrete symmetry based on the group $S_3$ might be an
adequate ingredient chosen by Nature. We have shown that both 
components arise naturally in a scheme with a replicated gauge group.

\section*{Acknowledgments}
This work is supported by the United States Department of Energy
under grant DE-FG02-97ER41029. The author would like to thank
the conference organizers for their hospitality.

\end{document}